\documentclass[sigconf]{acmart}

\usepackage{arydshln}

\hypersetup{
    colorlinks=true,
    linkcolor=blue,
    filecolor=magenta,      
    urlcolor=cyan,
    }

\usepackage[utf8]{inputenc} 
\usepackage[T1]{fontenc}    
\usepackage{hyperref}       
\usepackage{url}            
\usepackage{booktabs}       
\usepackage{amsfonts}       
\usepackage{nicefrac}       
\usepackage{microtype}      
\usepackage{xcolor}         

\usepackage{multirow}
\usepackage{enumitem}
\usepackage{adjustbox}
\usepackage{kotex}
\usepackage{rotating}
\usepackage{amsmath}
\usepackage{pifont}
\usepackage{array}
\usepackage{bbm}
\usepackage{wrapfig}
\usepackage[linesnumbered,algoruled,boxed,lined]{algorithm2e}
\usepackage{multicol}
\usepackage{caption}

\usepackage{lipsum}
\usepackage{graphicx}
\usepackage{float}

\newcommand{\cmark}{\ding{51}}
\newcommand{\xmark}{\ding{55}}

\newcommand{\ie}{{\it i.e.}}
\newcommand{\eg}{{\it e.g.}}

\newcommand{\topN}{top-\emph{N}}
\newcommand{\ul}{\underline}{}

\newcolumntype{P}[1]{>{\centering\arraybackslash}p{#1}}

\DeclareMathOperator*{\argmin}{argmin}

\title{Linear Item-Item Model with Neural Knowledge for Session-based Recommendation}

\author{Minjin Choi}
\affiliation{
  \institution{Samsung Research}
  \city{Seoul}
  \country{Republic of Korea}}
\email{min_jin.choi@samsung.com}

\author{Sunkyung Lee}
\affiliation{
  \institution{Sungkyunkwan University}
  \city{Suwon}
  \country{Republic of Korea}}
\email{sk1027@skku.edu}

\author{Seongmin Park}
\affiliation{
  \institution{Sungkyunkwan University}
  \city{Suwon}
  \country{Republic of Korea}}
\email{psm1206@skku.edu}

\author{Jongwuk Lee}\authornote{Corresponding author}
\affiliation{
  \institution{Sungkyunkwan University}
  \city{Suwon}
  \country{Republic of Korea}}
\email{jongwuklee@skku.edu}



\copyrightyear{2025}
\acmYear{2025}
\setcopyright{cc}
\setcctype{by-nc}
\acmConference[SIGIR '25]{Proceedings of the 48th International ACM SIGIR Conference on Research and Development in Information Retrieval}{July 13--18, 2025}{Padua, Italy}
\acmBooktitle{Proceedings of the 48th International ACM SIGIR Conference on Research and Development in Information Retrieval (SIGIR '25), July 13--18, 2025, Padua, Italy}
\acmDOI{10.1145/3726302.3730024}
\acmISBN{979-8-4007-1592-1/2025/07}

\begin{document}

\begin{abstract}

Session-based recommendation (SBR) aims to predict users' subsequent actions by modeling short-term interactions within sessions. Existing neural models primarily focus on capturing complex dependencies for sequential item transitions. As an alternative solution, linear item-item models mainly identify strong co-occurrence patterns across items and support faster inference speed. Although each paradigm has been actively studied in SBR, their fundamental differences in capturing item relationships and how to bridge these distinct modeling paradigms effectively remain unexplored. In this paper, we propose a novel SBR model, namely \emph{\textbf{L}inear \textbf{I}tem-Item model with \textbf{N}eural \textbf{K}nowledge (\textbf{LINK})}, which integrates both types of knowledge into a unified linear framework. Specifically, we design two specialized components of LINK: (i) \emph{Linear knowledge-enhanced Item-item Similarity model (LIS)}, which refines the item similarity correlation via self-distillation, and (ii) \emph{Neural knowledge-enhanced Item-item Transition model (NIT)}, which seamlessly incorporates complicated neural knowledge distilled from the off-the-shelf neural model. Extensive experiments demonstrate that LINK outperforms state-of-the-art linear SBR models across six real-world datasets, achieving improvements of up to 14.78\% and 11.04\% in Recall@20 and MRR@20 while showing up to 813x fewer inference FLOPs. Our code is available at \url{https://github.com/jin530/LINK}.

\end{abstract}

\maketitle

\section{Introduction}\label{sec:introduction}

Session-based recommendation (SBR) aims to predict the next item a user will likely consume by modeling short-term item interactions within an ongoing session. A session refers to a sequence of temporally correlated item interactions, \eg, sequential product clicks during an e-commerce transaction. Unlike traditional recommender models that leverage long-term user preferences and historical behaviors, SBR focuses on capturing immediate user interests through recent item sequences. It is particularly effective for anonymous users and cold-start scenarios in real-world applications, including e-commerce platforms, streaming services, and online advertising systems (\eg, Amazon, Spotify, and YouTube~\cite{LindenSY03Amazon, CovingtonAS16, ChenLSZ18Spotify}).

Capturing complex relationships between items is crucial for making appropriate recommendations. These relationships can be understood from two perspectives: item co-occurrence patterns (\ie, similar items appear together within sessions) and sequential transition patterns (\ie, sequential dependencies between consecutive items). To capture user interest in a session, most existing SBR models employ complex deep neural networks, including recurrent neural networks (RNNs)~\cite{HidasiKBT15GRU4Rec, LiRCRLM17NARM, HidasiK18GRU4Rec+}, graph neural networks (GNNs)~\cite{WuT0WXT19SRGNN, GuptaGMVS19NISER, PanCCCR20SGNNHN, ChenW20LESSR, Guo0SZWBZ22MSGIFSR, WangWCLMQ20GCEGNN, HuangCXXDCBZH21MTD, 0013YYWC021S2-DHCN, 0013YYSC21COTREC, ShiWL24SCL}, and Transformers~\cite{KangM18SASRec, YuanSSWZ21DSAN, XuZLSXZFZ19GCSAN, ChoKHY21ProxySR, LuoZLZWXFS20CoSAN, XieSLWGZDC22CL4SRec}. These neural models primarily focus on modeling sequential dependencies to capture complex relationships between items.

An alternative approach for SBR, recent studies~\cite{ChoiKLSL21SLIST, ChoiKLSL22S-Walk} have introduced linear item-item models~\cite{NingK11SLIM, Steck19EASE, Steck20EDLAE}. They effectively capture co-occurrence patterns between items by learning a single item-item similarity matrix, revealing which items tend to appear together within sessions. Notably, their simple structure enables efficient inference through a single matrix multiplication, making them particularly attractive for real-world applications. Although some studies attempt to incorporate sequential information through data augmentation~\cite{ChoiKLSL21SLIST} or random walk techniques~\cite{ChoiKLSL22S-Walk}, they still struggle to capture dynamic transitions that depend on the entire session context or complex sequential patterns. 
\begin{figure*}
\centering
\includegraphics[width=0.89\linewidth]{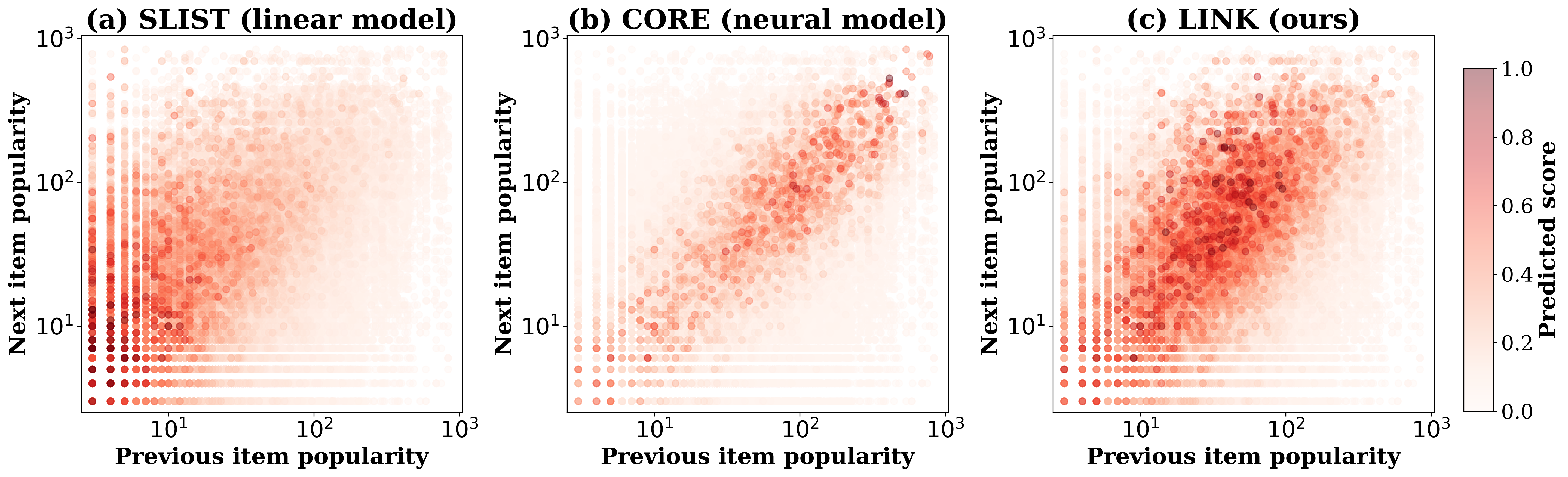}
\vspace{-3mm}
\caption{Comparison of how linear and neural models learn item relationships. The analysis is performed on consecutive item pairs (previous item $\rightarrow$ next item) from validation sessions in the Diginetica dataset. The x- and y-axis indicate the popularity of the previous and next items, respectively. The color represents the model's predicted correlation score. For simplicity, we linearly scale the predicted score into $[0, 1]$. }\label{fig:co_occur}


\vspace{-2.0mm}
\end{figure*}

%


Despite the widespread adoption of neural models in SBR, we question whether linear models might capture the patterns that neural models overlook, particularly given their fundamentally different architectural designs and learning mechanisms. While previous studies have analyzed neural and non-neural SBR models~\cite{LudewigMLJ19aneuralnon-neura, LudewigMLJ19bEmpiricalAnalysis} or experimented with simple ensemble methods~\cite{JannachL17rnnknn}, they have primarily focused on performance comparisons rather than understanding the fundamental differences in the knowledge captured by each paradigm. This lack of deep understanding has hindered the development of systematic approaches to integrating these distinct modeling paradigms, which could lead to enhanced modeling capabilities. To address this challenge, we raise two key questions: (i) \textit{How does the knowledge of neural and linear models differ in capturing item relationships?} (ii) \textit{How can we effectively bridge and integrate these different types of knowledge?}

We conduct an empirical analysis to understand how linear and neural models capture different types of item relationships. As shown in Figure~\ref{fig:co_occur}, our analysis reveals distinct patterns between the representative linear model (\ie, SLIST~\cite{ChoiKLSL21SLIST}) and neural model (\ie, CORE~\cite{hou2022core}). In Figure~\ref{fig:co_occur}(a), the linear model effectively captures strong collaborative signals between less popular items. By obtaining analytical solutions, linear models can learn salient correlation patterns even with limited interactions. In contrast, Figure~\ref{fig:co_occur}(b) demonstrates that the neural model excels at modeling relationships between relatively popular items, where abundant training examples allow neural models to leverage their expressive power to capture patterns that simpler linear models might miss. Inspired by this observation, we combine the two pieces of knowledge into a unified linear framework, enabling fast inference speed. As shown in Figure~\ref{fig:co_occur}(c), our proposed linear model improves upon existing linear models by distilling neural knowledge.

However, it is non-trivial to directly apply conventional knowledge distillation techniques~\cite{HintonVD15distillation, GouYMT21, FurlanelloLTIA18Bornagain} due to the architectural difference between linear and neural models. The key challenges arise from two aspects. (i) Traditional knowledge distillation assumes architectural similarity between teacher and student models, where both models process identical inputs and generate outputs in the same format. However, neural SBR models process variable-length sequences and learn complex embedding spaces, while linear models operate on fixed-size vectors and directly optimize an item-item matrix. (ii) Neural and linear models are trained with dissimilar learning objectives; neural models perform the next-item prediction task, while linear models optimize session vector reconstruction. These disparate objectives make it difficult to directly transfer different knowledge. To solve this dilemma, we propose a simple yet effective distillation method that transforms neural model predictions into linear transition matrices.

To this end, we propose \emph{\textbf{L}inear \textbf{I}tem-Item model with \textbf{N}eural \textbf{K}nowledge (\textbf{LINK})}, a novel framework that integrates the strengths of linear and neural models. Based on our empirical analysis, we design two specialized components to capture different aspects of item relationships, which are observed to be closely tied to item popularity. Co-occurrence patterns (items appearing together within a session) are structurally simpler and occur more frequently than transition patterns (items appearing in a specific sequential order), even for less popular items. This aligns well with linear models' ability to capture patterns from sparse interactions, making them particularly suited for modeling co-occurrence relationships. Meanwhile, neural models capture complex patterns, such as transitions, with abundant data from more popular items.

Specifically, LINK consists of \emph{Linear knowledge-enhanced Item-item Similarity model (LIS)} and \emph{Neural knowledge-enhanced Item-item Transition model (NIT)}. (i) LIS utilizes a self-distillation approach, allowing the linear model to act as its teacher. By leveraging the trained linear model, LIS refines the item-item similarity matrix to extend beyond intra-session similarity and capture broader inter-session relationships. (ii) NIT introduces a novel model-agnostic approach for incorporating neural knowledge into the linear framework. We develop a simple yet effective method that extracts an item-item transition matrix from the off-the-shelf neural model by aggregating its predictions on single-item sessions. By utilizing this matrix as a regularization term for the linear model, we can distill sequential knowledge from any neural architecture into the linear framework. This model-agnostic nature ensures that LINK can evolve with advances in neural architectures. Finally, we unify two components to capture both similarity and transition patterns, maintaining the computational efficiency of linear models.

In summary, the key contributions of this paper are as follows. 
\begin{itemize}[leftmargin=*,topsep=0pt,itemsep=-1ex,partopsep=1ex,parsep=1ex]

\vspace{0.5mm}
\item \textbf{Novel hybrid approach}: LINK effectively combines complementary knowledge from linear and neural models, as depicted in Figure~\ref{fig:co_occur}-(c). This enables LINK to capture complex item relationships that are challenging for a single modeling paradigm to learn. Moreover, its model-agnostic characteristic ensures that it evolves with advancements in neural architecture. 

\vspace{0.5mm}
\item \textbf{Novel knowledge transfer}: LINK introduces innovative techniques to bridge the architectural gap between linear and neural models, including self-distillation for similarity patterns and neural knowledge-based regularization for transition patterns. This approach successfully preserves the strengths of both modeling paradigms while maintaining computational efficiency.

\vspace{0.5mm}
\item \textbf{Extensive evaluation}: We extensively evaluated LINK across six session datasets. Our results show that LINK consistently outperforms existing linear models by up to 14.78\% and 11.04\% in Recall@20 and MRR@20. LINK also achieves up to 813 times fewer FLOPs than existing neural models as linear models.
\end{itemize}

\section{Related Work}

\subsection{Session-based Recommendation}


\subsubsection{\textbf{Neural SBR Models}} Many SBR models have utilized neural-based encoders, such as RNNs, GNNs, and Transformers.

\noindent
\textbf{RNN-based models.} GRU4Rec~\cite{HidasiKBT15GRU4Rec, HidasiK18GRU4Rec+} utilized the gated recurrent units (GRUs) to model the sequential item correlations. Subsequently, NARM~\cite{LiRCRLM17NARM} enhanced GRU4Rec~\cite{HidasiKBT15GRU4Rec} by incorporating both long-term and short-term preferences, and STAMP~\cite{LiuZMZ18STAMP} adopted an attention mechanism to capture users' general interests.

\noindent
\textbf{GNN-based models.} GNNs have become increasingly popular for SBR, representing each session as a small graph of items. SR-GNN~\cite{WuT0WXT19SRGNN} employed gated graph neural networks to capture multi-hop transitions among the session items, and NISER+~\cite{GuptaGMVS19NISER} mitigated the popularity bias and convergence issues through L2 embedding normalization. GC-SAN~\cite{XuZLSXZFZ19GCSAN} and FGNN~\cite{QiuLHY19FGNN} further improved upon SR-GNN~\cite{WuT0WXT19SRGNN} by incorporating self-attention mechanisms and weighted graph attention networks, respectively. LESSR~\cite{ChenW20LESSR} integrated GRU to address multi-hop information loss in GNN. SGNN-HN~\cite{PanCCCR20SGNNHN} introduced a star node to summarize sessions. GCE-GNN~\cite{WangWCLMQ20GCEGNN} and MTD~\cite{HuangCXXDCBZH21MTD} handled inter-session relationships by leveraging the item transition in global graphs derived from all sessions. $S^2$-DHCN~\cite{0013YYWC021S2-DHCN} and COTREC~\cite{0013YYSC21COTREC} utilized an additional supervised learning task to improve model training. Recent advancements include MSGIFSR~\cite{Guo0SZWBZ22MSGIFSR} and Atten-Mixer~\cite{ZhangGLXKZXWK23Atten-Mixer}, which aim to capture multiple user intents through varied item-level granularity using GNNs or multi-level reasoning over attention mixture networks. SCL~\cite{ShiWL24SCL} introduced self-contrastive learning to existing session-based recommenders. 
%

\noindent
\textbf{Transformer-based models.} Transformers have been actively adopted for recommendation tasks due to the powerful representation capacity for sequential data. SASRec~\cite{KangM18SASRec} is the canonical work in this area, which utilized a transformer-based encoder and regarded the last item representation as a session vector. DSAN~\cite{YuanSSWZ21DSAN} refined this approach by learning target embeddings to emulate the next target item, thus inferring user interests more precisely. Most recently, CORE~\cite{hou2022core} unified the representation space of session embeddings and item embeddings.



\subsubsection{\textbf{Linear SBR Models}} 

The primary objective of linear item-item models~\cite{NingK11SLIM, Steck19EASE, Steck20EDLAE, SteckL21Higher} is to learn an item-item similarity matrix from a given user-item matrix. Linear models are beneficial to accelerate the inference since computing \topN\ recommended items is simply done by a single matrix multiplication. SLIM~\cite{NingK11SLIM} formulated a linear model with constraints ensuring non-negative entries and a zero diagonal in the item-item matrix. EASE$^{\text{R}}$~\cite{Steck19EASE} simplified this approach by retaining only the zero-diagonal constraint, eliminating the non-negativity and $\ell_1$-norm constraints. Notably, the training complexity of EASE$^{\text{R}}$~\cite{Steck19EASE} scales with the number of items, which is typically much smaller than the number of sessions, offering a computational advantage.

SLIST~\cite{ChoiKLSL21SLIST} is the most representative work that adopted linear item-item models for SBR. It effectively models item correlations and dependencies using linear models, achieving accuracy comparable to neural models while maintaining high efficiency. SWalk~\cite{ChoiKLSL22S-Walk} employed a random walk approach to capture inter-session relationships more effectively. While linear models offer advantages in terms of efficiency compared to neural models, improving their accuracy remains a challenging task. To address this, our approach leverages knowledge from neural models to enhance accuracy while preserving the inherent efficiency of linear models.



\subsection{Knowledge Distillation} 
Knowledge distillation (KD) transfers the learned knowledge from a typically larger and more complex teacher model to a smaller student model~\cite{HintonVD15distillation, GouYMT21}. This concept has been extended to self-distillation in Born-Again Networks~\cite{FurlanelloLTIA18Bornagain}, where a model is iteratively refines itself using its own predictions. Knowledge distillation has been successfully employed in recommendation systems to enhance the performance of student models without increasing their size~\cite{TangW18, LeeCLS19, KangHKY20, Chen0F0023, KangKLL0Y23}. Inspired by this success, recent studies have extended knowledge distillation to the sequential recommendation tasks~\cite{WangWLGM00FDM22TiMiRec, DuYZZ00LS23EMKD}. For instance, TiMiRec~\cite{WangWLGM00FDM22TiMiRec} utilizes multi-interest distributions as pseudo-labels to enrich the target-interest distribution. While knowledge distillation between neural models has been extensively studied, integrating neural model knowledge into linear models presents unique challenges due to the fundamental architectural differences between neural and linear models.

\section{Proposed Model}\label{sec:model}

\begin{figure*}
\centering
\includegraphics[width=0.95\linewidth]{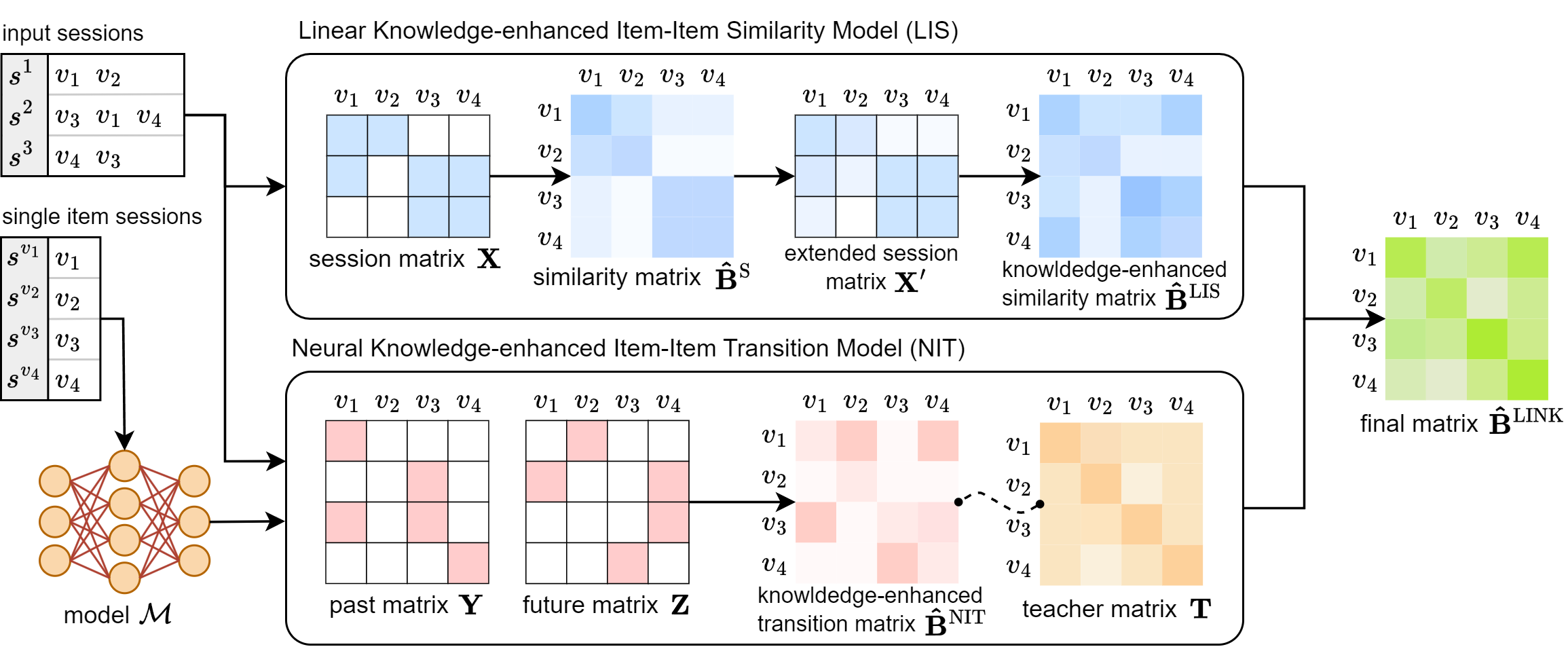}
\vspace{-3mm}
\caption{Overall training framework of LINK, consisting of Linear Knowledge-enhanced Item-Item Similarity Model (LIS) and Neural Knowledge-enhanced Item-Item Transition Model (NIT).}\label{fig:flow}
\Description{}{}
\vspace{-1mm}
\end{figure*}

In this section, we propose a \emph{\textbf{L}inear \textbf{I}tem-item model with \textbf{N}eural \textbf{K}nowledge (\textbf{LINK})}, a novel framework that effectively combines the complementary strengths of linear and neural models in session-based recommendation. LINK captures both co-occurrence and transition patterns by leveraging the unique advantages of each paradigm. Specifically, LINK offers the following benefits:
(i) While existing linear models focus primarily on direct co-occurrence patterns, LINK learns comprehensive item relationships by explicitly modeling both intra-session similarities and complex sequential transitions.
(ii) Through a novel knowledge distillation approach, LINK effectively incorporates sophisticated sequential patterns learned by neural models into its linear framework, enhancing its ability to capture dynamic transitions without sacrificing the inherent simplicity of linear models.
(iii) LINK presents a model-agnostic framework that can leverage knowledge from any state-of-the-art neural SBR model, allowing it to evolve with advances in neural architectures while maintaining its linear structure.

\vspace{0.5mm}
\noindent
\textbf{Overall framework.} LINK integrates linear and neural knowledge through two main components, as illustrated in Figure~\ref{fig:flow}: \emph{Linear knowledge-enhanced Item-item Similarity model (LIS)} and \emph{Neural knowledge-enhanced Item-item Transition model (NIT)}. LIS is designed to capture high-order co-occurrence patterns by leveraging self-distillation, where the model refines its own similarity matrix using its predictions as training samples (Section~\ref{sec:model_similarity}). On the other hand, NIT enhances sequential pattern modeling by incorporating transition knowledge from neural models, which is extracted through a novel approach of leveraging neural model responses to single-item sessions (Section~\ref{sec:model_transition}). Finally, we unify these two complementary perspectives by deriving a closed-form solution that combines both co-occurrence and transition patterns into a single item-item matrix (Section~\ref{sec:model_aggregation}).

\subsection{Linear Knowledge-enhanced Item-Item Similarity Model} \label{sec:model_similarity}

Let $\mathcal{S} = \{s^1, \dots, s^m\}$ denote a set of $m$ sessions, and $\mathcal{V} = \{v_1, \dots, v_n\}$ denote a set of $n$ unique items, \eg, products or songs.
An arbitrary session $s = [v_{t_1}, \dots, v_{t_{|s|}}] \in \mathcal{S}$ represents a sequence of $|s|$ items that a user interacts with, \eg, clicks, views, or purchases. Here, $t_i$ indicates the index of the $i$-th item in the session.
We represent session $s$ as a binary vector $\mathbf{x} \in \{0, 1\}^{n}$, where $\mathbf{x}_j = 1$ if $v_j$ is interacted with, and $\mathbf{x}_j = 0$ otherwise. By stacking $m$ sessions, let $\mathbf{X} \in \mathbb{R}^{m \times n}$ denote a session matrix.

We first introduce a linear model that captures the item co-occurrence within a session. It takes the session matrix $\mathbf{X} \in \mathbb{R}^{m \times n}$ as both an input and an output to obtain an item-item similarity matrix $\mathbf{B}^{\text{S}} \in \mathbb{R}^{n \times n}$~\cite{NingK11SLIM, Steck19EASE}. 
To avoid the trivial solution ($\hat{\mathbf{B}}^{\text{S}}=\mathbf{I}$) and to account for repeated items, we set constraints for the diagonal elements to be less than or equal to $\xi \in [0, 1)$~\cite{ChoiKLSL21SLIST}.
    \begin{equation}
    \label{eq:slis_loss}
    \hat{\mathbf{B}}^{\text{S}} = \argmin_{\mathbf{B}} \|\mathbf{X} - \mathbf{X} \cdot \mathbf{B}\|_F^2 + \lambda \|\mathbf{B}\|_F^2 \ \ \text{s.t.} \ \ \texttt{diag}(\mathbf{B}) \le \xi, 
    \end{equation}
where $\xi$ is the hyperparameter to control diagonal constraints, and $\lambda$ is the regularization coefficient.  
The closed-form solution of Eq.~\eqref{eq:slis_loss} is given below. (Please refer to \cite{ChoiKLSL21SLIST} for a detailed derivation.)
    \begin{equation}
        \hat{\mathbf{B}}^{\text{S}} = \mathbf{I} - \mathbf{P}  \cdot  \texttt{diagMat}(\mathbf{\gamma}),
        \ \ \mathbf{\gamma}_{j} = 
        \begin{cases}
            \ \lambda & \text{if} \ 1 - \lambda \mathbf{P}_{jj} \le \xi, \\
            \ \frac{1-\xi}{{\mathbf{P}}_{jj}} & \text{otherwise},
        \end{cases}
      \label{eq:slis_solution}
    \end{equation}
where $\mathbf{P} = (\mathbf{X}^{\top} \mathbf{X} + \lambda \mathbf{I})^{-1}$, $\texttt{diagMat}(\cdot)$ denotes the diagonal matrix, and $\mathbf{\gamma} \in \mathbb{R}^n$ is the vector for checking the diagonal constraint of $\hat{\mathbf{B}}^{\text{S}}$. In this way, the learned item-item similarity matrix $\hat{\mathbf{B}}^{\text{S}}$ captures the local similarity between items within a session, but the inter-session relationship is hardly captured.

To address this issue, we transform the session vector $\mathbf{x}$ with the knowledge of the trained linear model so that the training sample reflects the inter-session relationship. This is achieved by multiplying the original session vector $\mathbf{x}$ by the item-item similarity matrix $\hat{\mathbf{B}}^{\text{S}}$, which has learned intra-session relationship. It results in an extended session vector $\mathbf{x'}$ that reflects relevant items not present in the original session. This approach can also be viewed as a linear model variant of born-again networks~\cite{FurlanelloLTIA18Bornagain}, a knowledge distillation between the equal-sized teacher and student models.

We learn linear knowledge-enhanced item-item similarity matrix $\mathbf{B}^{\text{LIS}} \in \mathbb{R}^{n \times n}$ using extended session matrix $\mathbf{X}'$ as follows. 
    \begin{equation}
    \label{eq:sse_loss}
    \begin{split}
    \hat{\mathbf{B}}^{\text{LIS}} = \argmin_{\mathbf{B}} \|\mathbf{X}' - \mathbf{X}' \cdot \mathbf{B}\|_F^2 + & \lambda \|\mathbf{B}\|_F^2 \ \text{s.t.}\ \texttt{diag}(\mathbf{B}) \le \xi, \\
    \text{where}~\mathbf{X'}=\beta \mathbf{X}\hat{\mathbf{B}}^{\text{S}} 
    ~+~& (1 - \beta) \mathbf{X}.  
    \end{split}
    \end{equation}
Here, in forming an extended session matrix $\mathbf{X'}$, we introduce a hyperparameter $\beta$ that adjusts the ratio of the original intra-session and predicted inter-session relationship. The closed-form solution of Eq.~\eqref{eq:sse_loss} can be easily obtained by referring to Eq.~\eqref{eq:slis_solution}, \ie, we only need to change $\mathbf{P} = (\mathbf{X}^{\top} \mathbf{X} + \lambda \mathbf{I})^{-1}$ to $(\mathbf{X'}^{\top} \mathbf{X'} + \lambda \mathbf{I})^{-1}$.

The matrix multiplication operation $\mathbf{X'} = \mathbf{X}\hat{\mathbf{B}}^S$ ($\beta = 1$ for simplicity) enables our model to capture higher-order item relationships through information propagation across sessions, similar to how graph-based approaches identify multi-hop connections. To illustrate this mechanism, consider a simple example with three items $(i_1, i_2, i_3)$ in two sessions $s_1=(i_1, i_2)$ and $s_2=(i_2, i_3)$. While $i_1$ and $i_3$ do not directly co-occur in any session, multiplying the session vector with $\hat{B}^S$ reveals implicit relationships between these items through the shared item $i_2$. Specifically, $\hat{\mathbf{B}}^S$ captures first-order relationships ($i_1$$\leftrightarrow$$i_2$ and $i_2$$\leftrightarrow$$i_3$). When a session vector $x=(i_1,0,0)$ containing only item $i_1$ is multiplied by $\hat{B}^S$, the resulting $x'$ contains non-zero values not only for the directly connected $i_2$, but also for $i_3$ propagated via $i_2$. These inter-session relationships are then incorporated into $\hat{\mathbf{B}}^{\text{LIS}}$ through the optimization in Eq.~\eqref{eq:sse_loss} with a decay factor $\beta \in [0,1]$. This formulation allows us to balance between direct co-occurrences (original session vector $\mathbf{X}$) and higher-order relationships (propagated through $\hat{\mathbf{B}}^S$).

\subsection{Neural Knowledge-enhanced Item-Item Transition Model}\label{sec:model_transition}

Although LIS captures the similarity of items, it does not account for the sequential information within a session. To address this, we create two kinds of partial sessions based on each time step of the session $s$ to reflect the sequential order among the session items as in \cite{ChoiKLSL21SLIST, ChoiKLSL22S-Walk}: past partial session $s_{1:i-1} = [v_{t_{1}}, ..., v_{t_{i-1}}]$ and future partial session $s_{i:|s|} = [v_{t_{i}}, ..., v_{t_{|s|}}]$. Thus, we define $|s|-1$ pairs of past and future partial sessions for each session, \ie, $(s_{1:1}, s_{2:|s|}), \dots, (s_{1:|s-1|}, s_{|s|:|s|})$. Similar to converting session $s$ to a binary session vector $\mathbf{x}$ for learning item-item similarity models, we convert past and future partial sessions $s_{1:i-1}, s_{i:|s|}$ to session vectors $\mathbf{y}, \mathbf{z} \in \mathbb{R}^{n}$.

The importance of each item is determined by its distance from the last item in the past partial session (or the first item in the future partial session) using the following decay function.
    \begin{equation}
      \label{eq:w_position}
      w_\text{pos}(i, j, s) = \text{exp}\left(-\frac{|p(i, s) - p(j, s)|}{\delta_\text{pos}} \right),
    \end{equation}
where $\delta_\text{pos}$ is a hyperparameter controlling the magnitude of position decay, and $p(i, s)$ is the position of item $i$ in the session $s$. This decay function reduces the relevance between items $i$ and $j$ as their distance increases. For example, if $\delta_\text{pos}$ is 1, the $k$-th position away from the last item in the past session weights $\text{exp}(-k/1)$.

By stacking all partial session vectors, we construct past and future partial matrices $\mathbf{Y}, \mathbf{Z} \in \mathbb{R}^{m' \times n}$, where $m'$ is the total number of partial sessions. The item-item transition matrix $\mathbf{B}^{\text{T}} \in \mathbb{R}^{n \times n}$ is trained to recover the future partial matrix $\mathbf{Z}$ from the past partial matrix $\mathbf{Y}$, thereby capturing the sequential dependencies across items. Similar to Eq.~\eqref{eq:slis_loss}, we can easily derive the closed-form solution to the ridge regression problem.
\begin{equation}
\label{eq:slit_loss}
\begin{split}
\hat{\mathbf{B}}^{\text{T}} 
& = \argmin_{\mathbf{B}} \|\mathbf{Z} - \mathbf{Y} \cdot \mathbf{B}\|_F^2 + \lambda \|\mathbf{B}\|_F^2 \\
& = (\mathbf{Y}^{\top} \mathbf{Y} + \lambda \mathbf{I})^{-1}  \cdot (\mathbf{Y}^{\top} \mathbf{Z}).
\end{split}
\end{equation}
We enrich the item transition model with neural knowledge to leverage the sophisticated architecture and non-linearity of neural models. However, the distinct architecture and learning methods of neural and linear models make it challenging to obtain soft labels from the teacher model for training the student model, as is common in conventional knowledge distillation methods~\cite{HintonVD15distillation, WangWLGM00FDM22TiMiRec, KangKLL0Y23, angHKY21}.

To address this challenge, we introduce a novel method for directly extracting the item-item teacher matrix $\mathbf{T} \in \mathbb{R}^{n \times n}$ from the neural model. Specifically, a neural model $\mathcal{M}(\cdot)$ takes a session as input and outputs recommendation scores $\pi$ in logit form. Given a single item session $[v_i]$, we obtain the transition scores $\pi^{i}$ of all items from the item $v_i$. 
We extract each element of $\mathbf{T}$ as follows. 
    \begin{equation}
    \label{eq:teacher_single_item}
    \begin{split}
    \mathbf{T}_{i,j} & = \frac{\text{exp}(\pi^{i}_{j}/\tau)}{\sum_{k=1}^{n} \text{exp}(\pi^{i}_{k}/\tau)},  \ \ \text{where} \ 
    \pi^{i} = \mathcal{M}([v_i]) \ \text{for} \ i  \in \{1, \dots, n\},
    \end{split}
    \end{equation}
\noindent
where $\tau$ is a hyperparameter to adjust the temperature.

Since $\mathbf{T}$ is an item-item matrix, it can be employed as a linear model without further modification. However, as observed in Figure~\ref{fig:co_occur}, the knowledge from neural models and that of linear models are diverging. To seamlessly leverage both knowledge, we use $\mathbf{T}$ as regularization for learning the linear model. Since each row of $\mathbf{T}$ represents a probability distribution, we normalize the partial matrices $\mathbf{Y}$ and $\mathbf{Z}$ to convert them to probability values: $\mathbf{\Tilde{Z}} = \texttt{diagMat}(\mathbf{Z}\mathbf{1})^{-1}\mathbf{Z}$ and $\mathbf{\Tilde{Y}} = \texttt{diagMat}(\mathbf{Y}\mathbf{1})^{-1}\mathbf{Y}$, where $\mathbf{1} \in \mathbb{R}^{n}$ is ones-vector. 
To this end, we modify Eq.~\eqref{eq:slit_loss} as follows.
    \begin{equation}
    \label{eq:tke_loss}
    \hat{\mathbf{B}}^{\text{NIT}} = \argmin_{\mathbf{B}} \|\mathbf{\Tilde{Z}} - \mathbf{\Tilde{Y}} \cdot \mathbf{B}\|_F^2 + \lambda \|\textbf{T} - \mathbf{B}\|_F^2.
    \end{equation}

We revise the regularization term from $\|\mathbf{B}\|_F^2$ in Eq.~\eqref{eq:slit_loss} to $\|\textbf{T} - \mathbf{B}\|_F^2$, to incorporate knowledge from the neural model while learning linear relationship between items. We derive the following closed-form solution. 
    \begin{equation}
    \begin{split}
      \label{eq:itke_solution}
      \hat{\mathbf{B}}^{\text{NIT}} 
      & = (\mathbf{\Tilde{Y}}^{\top} \mathbf{\Tilde{Y}} + \lambda \mathbf{I})^{-1}  \cdot (\mathbf{\Tilde{Y}}^{\top} \mathbf{\Tilde{Z}} + \lambda \textbf{T} ).
    \end{split}
    \end{equation}


\subsection{Model Aggregation and Inference}\label{sec:model_aggregation}

\textbf{Model aggregation.}
To effectively exploit both item similarity and transition information, we combine Eq.~\eqref{eq:sse_loss} and Eq.~\eqref{eq:tke_loss} to obtain the final model $\mathbf{B}^{\text{LINK}} \in \mathbb{R}^{n \times n}$ as follows.
As $\mathbf{Y}$ and $\mathbf{Z}$ are normalized, we also normalize $\mathbf{X'}$ as $\mathbf{\Tilde{X}}' = \texttt{diagMat}(\mathbf{X'}\mathbf{1})^{-1}\mathbf{X'}$ to convert each row to a probability distribution. 
    \begin{align}
    \hat{\mathbf{B}}^{\text{LINK}}
    & = \argmin_{\mathbf{B}} \alpha \|\mathbf{\Tilde{X}}' - \mathbf{\Tilde{X}}' \cdot \mathbf{B}\|_F^2 + (1-\alpha) \|\mathbf{\Tilde{Z}} - \mathbf{\Tilde{Y}} \cdot \mathbf{B}\|_F^2 + \lambda \|\textbf{T} - \mathbf{B}\|_F^2 \nonumber \\
    & = \argmin_{\mathbf{B}} \|\mathbf{\Tilde{Z}}' - \mathbf{\Tilde{Y}}' \cdot \mathbf{B}\|_F^2 + \lambda \|\textbf{T} - \mathbf{B}\|_F^2, \\
    & \text{where}~\mathbf{\Tilde{Y}}'=\left[\begin{array}{r}
        \sqrt{\alpha}\mathbf{\Tilde{X}}' \\
        \sqrt{1\text{-}\alpha}\mathbf{\Tilde{Y}}
        \end{array}\right], ~\mathbf{\Tilde{Z}}'=\left[\begin{array}{r}
            \sqrt{\alpha}\mathbf{\Tilde{X}'} \\
            \sqrt{1\text{-}\alpha}\mathbf{\Tilde{Z}}
            \end{array}\right]   \nonumber \\
    & = (\mathbf{\Tilde{Y'}}^{\top} \mathbf{\Tilde{Y}'} + \lambda \mathbf{I})^{-1}  \cdot (\mathbf{\Tilde{Y}}^{\top} \mathbf{\Tilde{Z}'} + \lambda \textbf{T} ). \label{eq:link_solution}
    \end{align}
Here, $\alpha$ is the hyperparameter that controls the balance between LIS and NIT. Note that when $\alpha=0$, it is equal to Eq.~\eqref{eq:itke_solution}. 

\vspace{0.5mm}
\noindent
\textbf{Inference.} 
For a new session $s_\text{new}$, we construct a session vector $\mathbf{x}_\text{new} \in \mathbb{R}^{n}$. Each element of $\mathbf{x}_\text{new}$ is assigned an importance value based on the item's position in the session using the decay function $w_\text{inf}(i, s_{\text{new}})$ if the item is present in the session; otherwise, it is set to zero.
\begin{equation}
  \label{eq:weight_inference}
  w_\text{inf}(i, s_{\text{new}}) = \text{exp}\left(-\frac{|s_{\text{new}}| - p(i, s_{\text{new}})}{\delta_{\text{inf}}} \right),
\end{equation}
where $\delta_{\text{inf}}$ controls the position decay. For example, when $\delta_{\text{inf}}=1$, the last item of the new session weights $\text{exp}(-(0/1))$, and the second last item weights $\text{exp}(-(1/1))$. 
The final prediction for $s_{\text{new}}$ is obtained by multiplying the session vector and the item-item matrix: $\mathbf{x}_\text{new} \hat{\mathbf{B}}^{\text{LINK}}$. Note that $\mathbf{x}_\text{new}$ is a sparse vector with non-zero elements only for items in the session, requiring extremely low computational complexity during inference.

\subsection{Interpretation as Knowledge Distillation} \label{sec:interpret_kd}
We demonstrate how LIS (Eq.~\eqref{eq:sse_loss}) and NIT (Eq.~\eqref{eq:tke_loss}) can be understood through the framework of knowledge distillation~\cite{HintonVD15distillation, FurlanelloLTIA18Bornagain}, which transfer learned patterns from a teacher model to a student model through the following objective:
\begin{equation}
    \label{eq:KD}
    \mathcal{L}(f(x, \theta_{Teacher}), f(x, \theta_{Student})),
\end{equation}
where $\mathcal{L}$ represents a loss function (\eg, KL-divergence), and $f(x, \theta)$ denotes the model output given input $x$ and parameters $\theta$.

We show that both our key components, LIS and NIT, naturally align with this knowledge distillation framework. The objective function of LIS is interpreted as follows.
\begin{equation} \nonumber
    \label{eq:KD_LINK}
    \begin{split}
    \text{Eq.~\eqref{eq:sse_loss} objective} = & \|\mathbf{X}' - \mathbf{X}' \cdot \mathbf{B}\|_F^2 + \lambda \|\mathbf{B}\|_F^2 \\
    = & \|\mathbf{X}\hat{\mathbf{B}}^{\text{S}} - \mathbf{X}\hat{\mathbf{B}}^{\text{S}} \cdot \mathbf{B}\|_F^2 \text{\small ($\beta=0$ and $\lambda=0$ for simplicity)} \\
    = & \mathcal{L}(f(x, \theta_{Teacher}), f(x, \theta_{Student})),
    \end{split}
\end{equation}
\noindent
The loss function $\mathcal{L}$ is defined as the mean squared error using the Frobenius norm, and the function $f$ represents matrix multiplication. The teacher and student parameters are $\theta_{Teacher} = \hat{\mathbf{B}}^{\text{S}}$ and $\theta_{Student} = \hat{\mathbf{B}}^{\text{S}}\mathbf{B}$, respectively.

Similarly, the objective function of NIT is employed as follows.
\begin{equation} \nonumber
    \begin{split}
    \text{Eq.~\eqref{eq:tke_loss} objective} = & \|\mathbf{\Tilde{Z}} - \mathbf{\Tilde{Y}} \cdot \mathbf{B}\|_F^2 + \lambda \|\textbf{T} - \mathbf{B}\|_F^2 \\
    = & \|\mathbf{I}\textbf{T} - \mathbf{I}\mathbf{B}\|_F^2 \text{\small ($\lambda=1$ and omit non-distillation part)} \\
    = & \|\mathbf{X}\textbf{T} - \mathbf{X}\mathbf{B}\|_F^2 \text{\small ($\mathbf{X}$ equals sessions with single items)} \\
    = & \mathcal{L}(f(x, \theta_{Teacher}), f(x, \theta_{Student})),
    \end{split}
\end{equation}
\noindent
$\mathcal{L}$ and $f$ are mean squared error and matrix multiplication as before, with parameters $\theta_{Teacher}=\textbf{T}$ and $\theta_{Student}=\mathbf{B}$.

These formulations reveal that LIS performs self-distillation by using its predictions as targets, while NIT distills knowledge from a pre-trained neural model, both within the same mathematical framework of knowledge distillation.

\section{Experimental Setup}\label{sec:setup}



\textbf{Datasets.}
We conduct extensive experiments on six real-world session datasets collected from e-commerce and music streaming services: Diginetica\footnote{\url{https://competitions.codalab.org/competitions/11161}}, Retailrocket\footnote{\url{https://www.kaggle.com/retailrocket/ecommerce-dataset}}, Yoochoose\footnote{\url{https://www.kaggle.com/chadgostopp/recsys-challenge-2015}}, Tmall\footnote{\url{https://tianchi.aliyun.com/dataset/dataDetail?dataId=42}}, Dressipi\footnote{\url{https://dressipi.com/downloads/recsys-datasets/}}, and LastFM\footnote{\url{http://ocelma.net/MusicRecommendationDataset/lastfm-1K.html}}.
Additionally, we validate LINK with four datasets widely used in sequential recommendation (SR): Amazon\footnote{\url{https://cseweb.ucsd.edu/~jmcauley/datasets/amazon/links.html}} (Beauty, Toys, and Sports), and Yelp\footnote{\url{https://www.yelp.com/dataset/}}.
Table~\ref{tab:dataset} summarizes detailed statistics on all datasets. 
For data pre-processing, we discard the sessions/sequences with a single item and the items that occur less than five times in entire sessions/sequences to align with the conventional procedure~\cite{LudewigJ18Evaluation, LudewigMLJ19bEmpiricalAnalysis, LiRCRLM17NARM, WuT0WXT19SRGNN, DuYZQZ0LS23, DangYGJ0XSL23}. Following \cite{hou2022core}, we split training, validation, and test sets chronologically as the 8:1:1 ratio. 

\vspace{0.5mm}
\noindent
\textbf{Baselines.}
We compare the proposed model with representative SBR models, including two linear and fourteen neural models. 
For linear models, \textbf{SLIST}~\cite{ChoiKLSL21SLIST} is the first work that employs linear models designed for SBR. \textbf{SWalk}~\cite{ChoiKLSL22S-Walk} improves SLIST~\cite{ChoiKLSL21SLIST} by capturing the intra-session relationships via a random walk. 
For neural models, \textbf{GRU4Rec}~\cite{HidasiKBT15GRU4Rec, HidasiK18GRU4Rec+}, \textbf{NARM}~\cite{LiRCRLM17NARM} and \textbf{STAMP}~\cite{LiuZMZ18STAMP} utilize an RNN-based encoder for capturing the sequential interactions of users. \textbf{SASRec}~\cite{KangM18SASRec} uses a transformer-based encoder. \textbf{SR-GNN}~\cite{WuT0WXT19SRGNN} uses a GNN-based encoder to capture multi-hop item relationships. \textbf{NISER+}~\cite{GuptaGMVS19NISER} and \textbf{GC-SAN}~\cite{XuZLSXZFZ19GCSAN} enhance SR-GNN~\cite{WuT0WXT19SRGNN} via L2 normalization of item embedding vectors and a self-attention mechanism, respectively. \textbf{LESSR}~\cite{ChenW20LESSR} reduces multi-hop information loss in GNN via GRU. \textbf{GCE-GNN}~\cite{WangWCLMQ20GCEGNN} handles inter-session relationships by taking the item transition in the global graph from whole sessions. \textbf{SGNN-HN}~\cite{PanCCCR20SGNNHN} proposes a star graph neural network to model the complex transition. \textbf{DSAN}~\cite{YuanSSWZ21DSAN} applies a sparse transformation function to denoise the irrelevant items. \textbf{CORE}~\cite{hou2022core} is a transformer-based model representing sessions and items on a consistent embedding space. \textbf{MSGIFSR}~\cite{Guo0SZWBZ22MSGIFSR} adopts consecutive item units for representing different item-level granularity. Recently, \textbf{SCL}~\cite{ShiWL24SCL} enhances the contrastive learning task for SBR. We adopt the GCE-GNN version of SCL as the baseline. 
Furthermore, we compare the performance of LINK with three state-of-the-art models in sequential recommendation protocol. \textbf{DuoRec}~\cite{QiuHYW22} enhances SASRec~\cite{KangM18SASRec} by employing contrastive learning. \textbf{FEARec}~\cite{DuYZQZ0LS23} incorporates frequency knowledge to better capture user sequences. \textbf{BSARec}~\cite{ShinCWP24} utilizes an inductive bias to account for granular sequential patterns.

\begin{table}[] \small
\centering
\caption{Statistics of the various benchmark datasets. AvgLen indicates the average length of entire sessions/sequences.}\label{tab:dataset}
\vspace{-3mm}
\setlength{\tabcolsep}{8.3pt} 
\begin{tabular}{c|cccc}
\toprule
Dataset      & \# Interacts & \# Sessions & \# Items & AvgLen \\
\midrule
Diginetica   & 786,582         & 204,532     & 42,862   & 4.12        \\
Retailrocket & 871,637         & 321,032     & 51,428   & 6.40        \\
Yoochoose    & 1,434,349       & 470,477     & 19,690   & 4.64        \\
Tmall        & 427,797         & 66,909      & 37,367   & 10.62       \\
Dressipi     & 4,305,641       & 943,658     & 18,059   & 6.47        \\
LastFM       & 3,510,163       & 325,543     & 38,616   & 8.16        \\
\midrule
Beauty       & 198,502         & 22,363      & 12,101   & 8.9         \\
Toys         & 167,597         & 19,412      & 11,924   & 8.6         \\
Sports       & 296,337         & 29,858      & 18,357   & 8.3         \\
Yelp         & 317,078         & 30,494      & 20,061   & 10.4        \\
\bottomrule
\end{tabular}
\vspace{-2mm}
\end{table}

\vspace{0.5mm}
\noindent
\textbf{Evaluation protocol and metrics.}
Following the common protocol to evaluate SBR models~\cite{LiRCRLM17NARM, WuT0WXT19SRGNN}, we adopt the \emph{iterative revealing scheme}, which iteratively exposes items from a session to the model. We adopt Recall (R@20) and Mean Reciprocal Rank (M@20) to quantify the prediction accuracy for the next item. For comparison with SR models, we employ the \emph{leave-one-out}~\cite{DuYZQZ0LS23, DangYGJ0XSL23, ShinCWP24} protocol and full ranking scenario. We additionally employ Normalized Discounted Cumulative Gain (N@20) for SR models. 

\begin{table*} \small
\centering
\setlength{\tabcolsep}{3.2pt} 
\caption{Performance comparison of SBR models, including standard deviations for neural models. Gain indicates how much better LINK is than the best linear model. The best model is marked in \textbf{bold}, and the second-best model is \ul{underlined}. All experimental results for neural models are averaged over five runs using different random seeds. Note that linear models, \eg, SLIST~\cite{ChoiKLSL21SLIST}, SWalk~\cite{ChoiKLSL22S-Walk} and LINK, exhibit no variance as they are deterministic.}\label{tab:exp_all}
\vspace{-3.5mm}
\begin{tabular}{c|cc|cc|cc|cc|cc|cc}
\toprule
\multicolumn{1}{c|}{\multirow{2}{*}{Model}} & \multicolumn{2}{c|}{Diginetica} & \multicolumn{2}{c|}{Retailrocket} & \multicolumn{2}{c|}{Yoochoose} & \multicolumn{2}{c|}{Tmall} & \multicolumn{2}{c|}{Dressipi} & \multicolumn{2}{c}{LastFM} \\
\multicolumn{1}{c|}{} & R@20 & \multicolumn{1}{c|}{M@20} & R@20 & \multicolumn{1}{c|}{M@20} & R@20 & \multicolumn{1}{c|}{M@20} & R@20 & \multicolumn{1}{c|}{M@20} & R@20 & \multicolumn{1}{c|}{M@20} & R@20 & M@20 \\ \midrule
\multicolumn{13}{c}{\textit{Neural Models}} \\ \midrule
\multicolumn{1}{c|}{GRU4Rec~\cite{HidasiK18GRU4Rec+}} & 47.19\scriptsize$\pm$0.19 & \multicolumn{1}{c|}{15.73\scriptsize$\pm$0.04} & 57.79\scriptsize$\pm$0.07 & \multicolumn{1}{c|}{34.87\scriptsize$\pm$0.13} & 61.16\scriptsize$\pm$0.26 & 27.19\scriptsize$\pm$0.15 & 33.87\scriptsize$\pm$0.15 & \multicolumn{1}{c|}{23.37\scriptsize$\pm$0.28} & 35.33\scriptsize$\pm$0.31 & \multicolumn{1}{c|}{13.74\scriptsize$\pm$0.14} & 20.88\scriptsize$\pm$0.11 & 7.21\scriptsize$\pm$0.07 \\
\multicolumn{1}{c|}{NARM~\cite{LiRCRLM17NARM}} & 48.09\scriptsize$\pm$0.18 & \multicolumn{1}{c|}{16.01\scriptsize$\pm$0.09} & 59.21\scriptsize$\pm$0.18 & \multicolumn{1}{c|}{35.61\scriptsize$\pm$0.12} & 62.43\scriptsize$\pm$0.21 & 28.25\scriptsize$\pm$0.12 & 35.55\scriptsize$\pm$0.27 & \multicolumn{1}{c|}{24.58\scriptsize$\pm$0.18} & 37.40\scriptsize$\pm$0.21 & \multicolumn{1}{c|}{14.80\scriptsize$\pm$0.08} & 20.91\scriptsize$\pm$0.19 & 7.34\scriptsize$\pm$0.03 \\
\multicolumn{1}{c|}{STAMP~\cite{LiuZMZ18STAMP}} & 44.25\scriptsize$\pm$0.10 & \multicolumn{1}{c|}{14.84\scriptsize$\pm$0.05} & 53.86\scriptsize$\pm$0.18 & \multicolumn{1}{c|}{31.89\scriptsize$\pm$0.14} & 60.50\scriptsize$\pm$0.33 & 27.87\scriptsize$\pm$0.10 & 22.79\scriptsize$\pm$2.10 & \multicolumn{1}{c|}{15.67\scriptsize$\pm$1.11} & 36.17\scriptsize$\pm$0.23 & \multicolumn{1}{c|}{14.56\scriptsize$\pm$0.14} & 19.93\scriptsize$\pm$0.29 & 7.40\scriptsize$\pm$0.07 \\
\multicolumn{1}{c|}{SASRec~\cite{KangM18SASRec}} & 35.73\scriptsize$\pm$7.13 & \multicolumn{1}{c|}{13.14\scriptsize$\pm$2.10} & 58.93\scriptsize$\pm$0.13 & \multicolumn{1}{c|}{35.32\scriptsize$\pm$0.08} & 62.47\scriptsize$\pm$0.35 & 27.77\scriptsize$\pm$0.14 & 36.32\scriptsize$\pm$0.16 & \multicolumn{1}{c|}{25.66\scriptsize$\pm$0.10} & 32.71\scriptsize$\pm$0.39 & \multicolumn{1}{c|}{12.28\scriptsize$\pm$0.23} & 14.49\scriptsize$\pm$0.77 & 4.15\scriptsize$\pm$0.29 \\
\multicolumn{1}{c|}{SR-GNN~\cite{WuT0WXT19SRGNN}} & 48.41\scriptsize$\pm$0.16 & \multicolumn{1}{c|}{16.45\scriptsize$\pm$0.12} & 58.72\scriptsize$\pm$0.10 & \multicolumn{1}{c|}{35.49\scriptsize$\pm$0.07} & 62.10\scriptsize$\pm$0.32 & 28.17\scriptsize$\pm$0.11 & 34.78\scriptsize$\pm$0.31 & \multicolumn{1}{c|}{24.82\scriptsize$\pm$0.49} & 35.62\scriptsize$\pm$0.23 & \multicolumn{1}{c|}{14.14\scriptsize$\pm$0.16} & 21.45\scriptsize$\pm$0.20 & 8.63\scriptsize$\pm$0.08 \\
\multicolumn{1}{c|}{NISER+~\cite{GuptaGMVS19NISER}} & 50.80\scriptsize$\pm$0.25 & \multicolumn{1}{c|}{17.91\scriptsize$\pm$0.16} & 60.70\scriptsize$\pm$0.12 & \multicolumn{1}{c|}{37.95\scriptsize$\pm$0.12} & 63.33\scriptsize$\pm$0.26 & 29.02\scriptsize$\pm$0.08 & 40.17\scriptsize$\pm$0.10 & \multicolumn{1}{c|}{29.16\scriptsize$\pm$0.04} & 37.97\scriptsize$\pm$0.10 & \multicolumn{1}{c|}{15.13\scriptsize$\pm$0.10} & 22.15\scriptsize$\pm$0.09 & \ul{8.64}\scriptsize$\pm$0.02 \\
\multicolumn{1}{c|}{GC-SAN~\cite{XuZLSXZFZ19GCSAN}} & 49.94\scriptsize$\pm$0.17 & \multicolumn{1}{c|}{17.39\scriptsize$\pm$0.08} & 60.04\scriptsize$\pm$0.04 & \multicolumn{1}{c|}{36.23\scriptsize$\pm$0.11} & 62.70\scriptsize$\pm$0.33 & 28.68\scriptsize$\pm$0.14 & 36.60\scriptsize$\pm$0.08 & \multicolumn{1}{c|}{24.65\scriptsize$\pm$0.09} & 37.62\scriptsize$\pm$0.12 & \multicolumn{1}{c|}{14.97\scriptsize$\pm$0.03} & 21.83\scriptsize$\pm$0.11 & 8.54\scriptsize$\pm$0.02 \\
\multicolumn{1}{c|}{LESSR~\cite{ChenW20LESSR}} & 48.45\scriptsize$\pm$0.16 & \multicolumn{1}{c|}{16.33\scriptsize$\pm$0.21} & 57.75\scriptsize$\pm$1.18 & \multicolumn{1}{c|}{36.35\scriptsize$\pm$0.20} & 62.51\scriptsize$\pm$0.45 & 28.25\scriptsize$\pm$0.21 & 33.97\scriptsize$\pm$0.22 & \multicolumn{1}{c|}{24.40\scriptsize$\pm$0.18} & 36.84\scriptsize$\pm$0.29 & \multicolumn{1}{c|}{14.30\scriptsize$\pm$0.06} & 21.93\scriptsize$\pm$0.15 & 8.56\scriptsize$\pm$0.06 \\
\multicolumn{1}{c|}{GCE-GNN~\cite{WangWCLMQ20GCEGNN}} & 50.46\scriptsize$\pm$0.06 & \multicolumn{1}{c|}{17.71\scriptsize$\pm$0.09} & 60.42\scriptsize$\pm$0.16 & \multicolumn{1}{c|}{35.70\scriptsize$\pm$0.05} & 63.31\scriptsize$\pm$0.13 & 28.95\scriptsize$\pm$0.13 & 36.25\scriptsize$\pm$0.21 & \multicolumn{1}{c|}{26.73\scriptsize$\pm$0.13} & 37.23\scriptsize$\pm$0.24 & \multicolumn{1}{c|}{14.97\scriptsize$\pm$0.06} & 22.75\scriptsize$\pm$0.13 & 8.35\scriptsize$\pm$0.03 \\
\multicolumn{1}{c|}{SGNN-HN~\cite{PanCCCR20SGNNHN}} & 50.58\scriptsize$\pm$0.10 & \multicolumn{1}{c|}{17.10\scriptsize$\pm$0.03} & 58.64\scriptsize$\pm$0.57 & \multicolumn{1}{c|}{35.31\scriptsize$\pm$0.09} & 61.99\scriptsize$\pm$0.31 & 28.23\scriptsize$\pm$0.05 & 39.66\scriptsize$\pm$0.33 & \multicolumn{1}{c|}{23.62\scriptsize$\pm$0.16} & 38.39\scriptsize$\pm$0.18 & \multicolumn{1}{c|}{15.05\scriptsize$\pm$0.08} & 22.72\scriptsize$\pm$0.10 & 7.72\scriptsize$\pm$0.04 \\
\multicolumn{1}{c|}{DSAN~\cite{YuanSSWZ21DSAN}} & 51.64\scriptsize$\pm$0.17 & \multicolumn{1}{c|}{18.32\scriptsize$\pm$0.04} & 61.00\scriptsize$\pm$0.48 & \multicolumn{1}{c|}{38.76\scriptsize$\pm$0.06} & 63.44\scriptsize$\pm$0.18 & \ul{29.08}\scriptsize$\pm$0.14 & 42.87\scriptsize$\pm$0.25 & \multicolumn{1}{c|}{30.94\scriptsize$\pm$0.04} & 37.19\scriptsize$\pm$0.14 & \multicolumn{1}{c|}{14.86\scriptsize$\pm$0.07} & 22.49\scriptsize$\pm$0.09 & 7.92\scriptsize$\pm$0.06 \\
\multicolumn{1}{c|}{CORE~\cite{hou2022core}} & \ul{52.87}\scriptsize$\pm$0.09 & \multicolumn{1}{c|}{18.59\scriptsize$\pm$0.05} & 62.00\scriptsize$\pm$0.07 & \multicolumn{1}{c|}{38.17\scriptsize$\pm$0.13} & 64.57\scriptsize$\pm$0.05 & 28.35\scriptsize$\pm$0.02 & 44.88\scriptsize$\pm$0.10 & \multicolumn{1}{c|}{\ul{31.49}\scriptsize$\pm$0.04} & 38.33\scriptsize$\pm$0.13 & \multicolumn{1}{c|}{15.70\scriptsize$\pm$0.03} & 22.82\scriptsize$\pm$0.13 & 8.16\scriptsize$\pm$0.04 \\
\multicolumn{1}{c|}{MSGIFSR~\cite{Guo0SZWBZ22MSGIFSR}} & 52.54\scriptsize$\pm$0.16 & \multicolumn{1}{c|}{\ul{19.06}\scriptsize$\pm$0.08} & \ul{62.67}\scriptsize$\pm$0.11 & \multicolumn{1}{c|}{\textbf{39.08}\scriptsize$\pm$0.05} & \ul{64.81}\scriptsize$\pm$0.09 & \textbf{30.41}\scriptsize$\pm$0.05 & \ul{53.07}\scriptsize$\pm$0.05 & \multicolumn{1}{c|}{29.20\scriptsize$\pm$0.14} & \ul{40.33}\scriptsize$\pm$0.10 & \multicolumn{1}{c|}{16.11\scriptsize$\pm$0.06} & \ul{25.00}\scriptsize$\pm$0.10 & \textbf{10.17}\scriptsize$\pm$0.04 \\
\multicolumn{1}{c|}{SCL~\cite{ShiWL24SCL}} & 52.03\scriptsize$\pm$0.14 & \multicolumn{1}{c|}{18.84\scriptsize$\pm$0.09} & 60.12\scriptsize$\pm$0.11 & \multicolumn{1}{c|}{37.88\scriptsize$\pm$0.13} & 63.47\scriptsize$\pm$0.17 & \ul{29.07}\scriptsize$\pm$0.11 & 38.55\scriptsize$\pm$0.30 & \multicolumn{1}{c|}{28.33\scriptsize$\pm$0.13} & 38.08\scriptsize$\pm$0.07 & \multicolumn{1}{c|}{15.29\scriptsize$\pm$0.06} & 22.20\scriptsize$\pm$1.72 & 7.93\scriptsize$\pm$1.00 \\ \midrule
\multicolumn{13}{c}{\textit{Linear Models}} \\ \midrule
\multicolumn{1}{c|}{SLIST~\cite{ChoiKLSL21SLIST}} & 49.81 & \multicolumn{1}{c|}{18.45} & 60.10 & \multicolumn{1}{c|}{37.69} & 63.32 & \multicolumn{1}{c|}{28.30} & 47.43 & \multicolumn{1}{c|}{28.93} & 40.20 & \multicolumn{1}{c|}{16.13} & 24.63 & 7.73 \\
\multicolumn{1}{c|}{SWalk~\cite{ChoiKLSL22S-Walk}} & 51.19 & \multicolumn{1}{c|}{18.76} & 61.31 & \multicolumn{1}{c|}{38.79} & 63.72 & \multicolumn{1}{c|}{27.61} & 48.77 & \multicolumn{1}{c|}{29.52} & 39.80 & \multicolumn{1}{c|}{\ul{16.19}} & 24.84 & 7.70 \\ \midrule
\multicolumn{1}{c|}{LINK (Ours)} & \textbf{53.62} & \multicolumn{1}{c|}{\textbf{19.22}} &  \textbf{62.74} & \multicolumn{1}{c|}{\ul{38.97}} & \textbf{64.90} & \multicolumn{1}{c|}{28.82} &  \textbf{55.98} & \multicolumn{1}{c|}{\textbf{32.78}} &  \textbf{41.39} & \multicolumn{1}{c|}{\textbf{16.38}} & \textbf{25.14} & 7.97 \\ \midrule
\multicolumn{1}{c|}{Gain (\%)} & 4.75 & \multicolumn{1}{c|}{2.45} & 2.33 & \multicolumn{1}{c|}{0.46} & 1.85 & \multicolumn{1}{c|}{1.84} & 14.78 & \multicolumn{1}{c|}{11.04} & 2.96 & \multicolumn{1}{c|}{1.17} & 1.21 & 3.10 \\ 
\bottomrule
\end{tabular}
\vspace{-2mm}
\end{table*}

\vspace{0.5mm}
\noindent
\textbf{Implementation details.}
We optimize all the baselines using Adam optimizer~\cite{KingmaB14Adam} with a learning rate of 0.001. We set the dimension of the embedding vectors to 100, the max session/sequence length to 50, and the batch size to 512. We stopped the training if the validation M@20 decreased for three consecutive epochs. We finally report the performance on the test set using the models that show the highest performance on the validation set. For LINK, we used CORE~\cite{hou2022core} as a default teacher model. We tuned $\alpha$ and $\beta$ from 0 to 1 in 0.1 increments, $\lambda$ among \{$10^{-1}$, $10^{0}$, $10^{1}$, $10^{2}$, $10^{3}$, $10^{4}$\}, and $\tau$ among \{0.01, 0.02, 0.05, 0.1, 0.2, 0.5, 1.0\}, respectively. We searched $\delta_{\text{pos}}$ and $\delta_{\text{inf}}$ from $2^{-3}$ to $2^{3}$ in 2x increments. 

For baseline models, we follow the original papers' hyperparameter settings and thoroughly tune them if they are unavailable. For neural models, all experimental results are averaged over five runs with different seeds. For reproducibility, we implement all baseline models on an open-source recommendation system library RecBole\footnote{\url{https://github.com/RUCAIBox/RecBole}}~\cite{ZhaoMHLCPLLWTMF21RecBole, ZhaoHPYZLZBTSCX22RecBole2.0}. For the efficiency study in Figure~\ref{fig:exp_flops_digi_yc}, we utilized the fvcore library\footnote{\url{https://github.com/facebookresearch/fvcore}} to measure the FLOPs of neural models. For MSGIFSR~\cite{Guo0SZWBZ22MSGIFSR} which is implemented using the dgl library\footnote{\url{https://github.com/dmlc/dgl/}}, the FLOPs could not be measured. Thus, we calculated the FLOPs of MSGIFSR~\cite{Guo0SZWBZ22MSGIFSR} by averaging the FLOPs of the GNN-based SBR models except SCL~\cite{ShiWL24SCL} and GCE-GNN~\cite{WangWCLMQ20GCEGNN}. 



\section{Experimental Results}\label{sec:results}

\subsection{Overall Evaluation}

\noindent
\textbf{Effectiveness in SBR.}
Table~\ref{tab:exp_all} reports the overall accuracy of LINK compared to competitive models across six datasets. The key observations are as follows. 
(i) Compared to state-of-the-art neural models, LINK shows comparable or superior performance, achieving up to 5.49\% and 4.08\% improvement in R@20 and M@20 over the most competitive neural models. This improvement is particularly notable as LINK effectively combines the strengths of both modeling paradigms.
(ii) Among linear models, LINK establishes new state-of-the-art performance, with improvements up to 14.78\% in R@20 and 11.04\% in M@20. It is shown that LINK has successfully distilled neural knowledge that was hardly captured by the linear model, \eg, sophisticated transition patterns between items.

\begin{figure*}
\begin{tabular}{cc}
\includegraphics[width=0.45\linewidth]{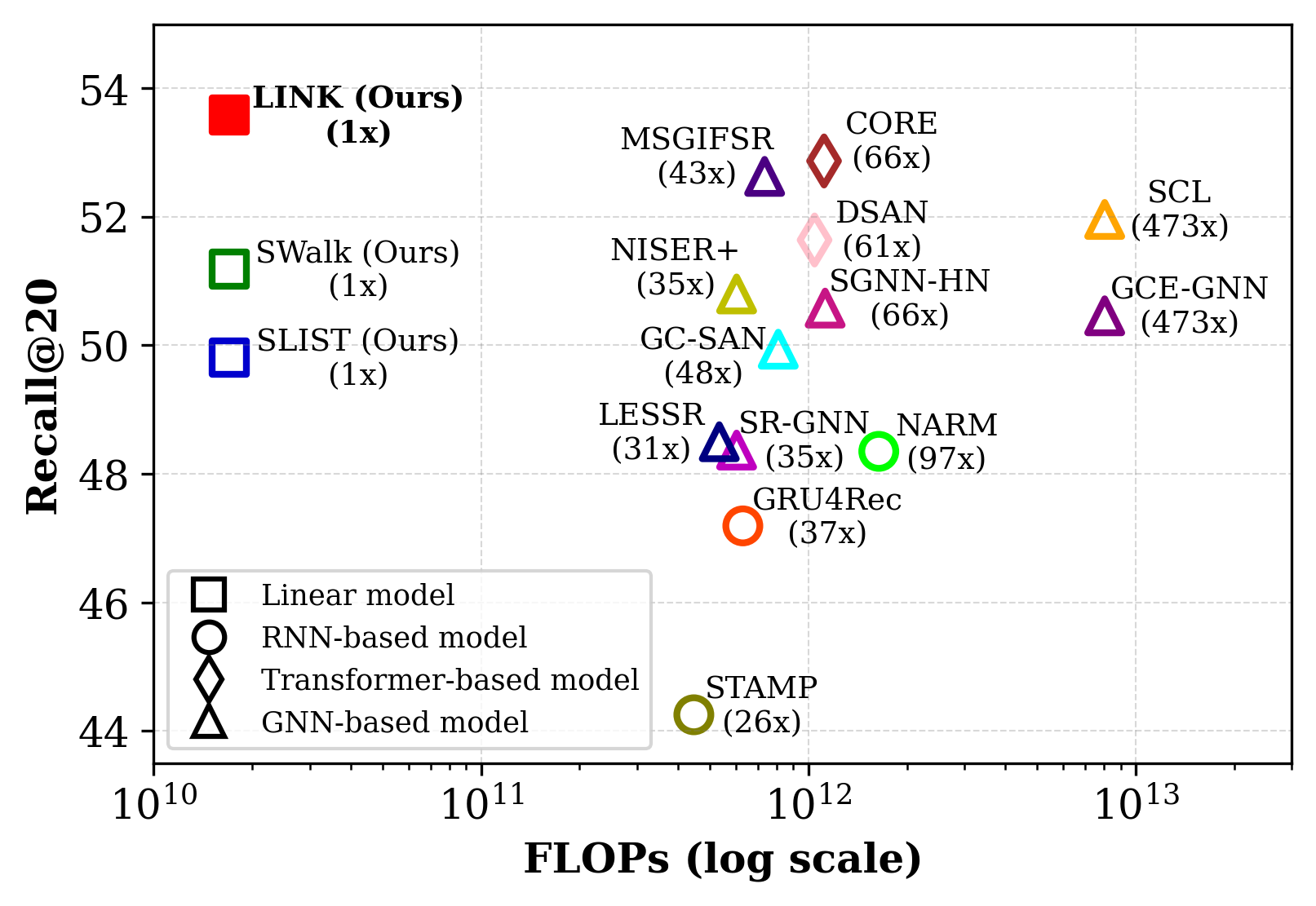} 
& \includegraphics[width=0.45\linewidth]{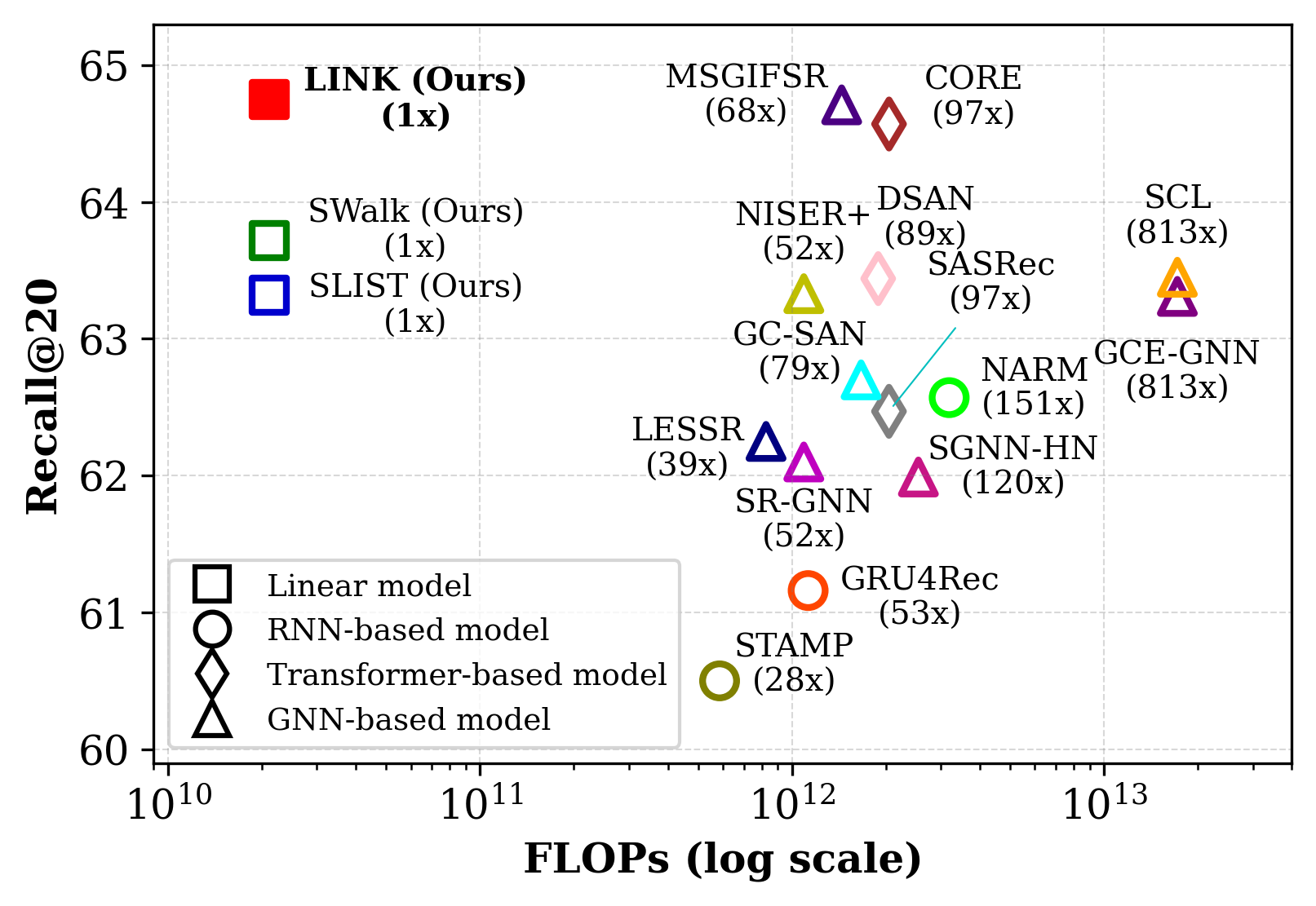} \\
(a) Diginetica & (b) Yoochoose \\
\end{tabular}
\vspace{-2mm}
\caption{Performance comparison over the Recall@20 and the log-scaled number of floating point operations (FLOPs) for inference on Diginetica and Yoochoose datasets. The numbers in parentheses indicate the relative FLOPs compared to LINK.}\label{fig:exp_flops_digi_yc}
\Description{}{}
\vspace{-1mm}
\end{figure*}

\vspace{0.5mm}
\noindent
\textbf{Inference efficiency in SBR.}
Figure~\ref{fig:exp_flops_digi_yc} presents the trade-off between the accuracy and efficiency of SBR models on Diginetica and Yoochoose datasets. 
(i) Linear models demonstrate significantly higher efficiency than neural models, requiring up to 813 times fewer flops on Yoochoose. The efficiency stems from the fact that inference in linear models requires only a single matrix multiplication to compute the top-$N$ item lists. This characteristic makes LINK well-suited to meet the practical efficiency constraints of commercial recommendation systems. 
(ii) Among linear models, LINK exhibits superior performance, elevating the accuracy limits of linear models while maintaining their computational efficiency. It indicates how LINK successfully mitigates the accuracy-efficiency trade-off compared to existing models.

\begin{table*} \small
\centering
\setlength{\tabcolsep}{11.8pt} 
\caption{Training times for SBR models. Numbers in parentheses indicate the relative training time compared to LINK. For LINK, we exclude the training time required for linear and neural teachers (\ie, $\mathbf{B}^{\text{S}}$ and $\mathcal{M}$).}\label{tab:training_time}
\vspace{-3.5mm}
\begin{tabular}{c|rrrrrr}
\toprule
\multirow{2}{*}{Model} & \multicolumn{1}{c}{Diginetica} & \multicolumn{1}{c}{Retailrocket} & \multicolumn{1}{c}{Yoochoose} & \multicolumn{1}{c}{Tmall} & \multicolumn{1}{c}{Dressipi} & \multicolumn{1}{c}{LastFM} \\
& \multicolumn{6}{c}{Overall training time (in seconds)} \\
\midrule
CORE~\cite{hou2022core} & 2,141s (4.4x) & 1,607s (2.0x) & 7,661s (55.3x) & 398s (1.3x) & 20,225s (221.6x) & 6,443s (19.0x) \\
MSGIFSR~\cite{Guo0SZWBZ22MSGIFSR} & 11,639s (23.9x) & 13,903s (17.1x) & 41,225s (297.7x) & 5,169s (17.0x) & 87,028s (953.6x) & 58,000s (170.8x) \\
SCL~\cite{ShiWL24SCL} & 3,690s (7.6x) & 3,480s (4.3x) & 8,589s (62.0x) & 571s (1.9x) & 19,816s (217.1x) & 15,986s (47.1x) \\
\midrule
SLIST~\cite{ChoiKLSL21SLIST} & 355s (0.7x) & 599s (0.7x) & 42s (0.3x) & 246s (0.8x) & 49s (0.5x) & 273s (0.8x) \\ 
SWalk~\cite{ChoiKLSL22S-Walk} & 2,021s (4.1x) & 3,272s (4.0x) & 259s (1.9x) & 970s (3.2x) & 237s (2.6x) & 1,398s (4.1x) \\
\midrule
LINK (Ours) & 487s (1.0x) & 814s (1.0x) & 138s (1.0x) & 305s (1.0x) & 91s (1.0x) & 340s (1.0x) \\ 
\bottomrule
\end{tabular}
\vspace{-2mm}
\end{table*}

\vspace{0.5mm}
\noindent
\textbf{Training efficiency in SBR.}
Table~\ref{tab:training_time} shows the overall training time of various SBR models. The key findings can be summarized as follows. 
(i) LINK generally demonstrates faster training times than existing methods, except SLIST. The training process of LINK can be divided into two main parts: knowledge extraction from linear and neural models (\ie, $\mathbf{X}'$ and $\mathbf{T}$) and computing the closed-form solution (\ie, Eq.~\eqref{eq:link_solution}). The latter requires a similar computation time as SLIST, and knowledge extraction accounts for 20--70\% of the total training time.
(ii) LINK significantly outperforms neural models in training speed, achieving up to 953x speedup on the Dressipi dataset. This significant improvement can be attributed to the computational complexity of the closed-form solution, which scales with the number of items rather than the number of sessions~\cite{Steck19EASE}. This characteristic enables LINK to maintain its efficiency advantage even as the dataset size increases.
(iii) While we report LINK's training time excluding teacher model training (as they can be pre-trained and reused), the total training remains efficient, including teacher models. For instance, even when including both CORE and SLIST training time, LINK achieves 4.7x faster training compared to MSGIFSR on the Tmall dataset.

\vspace{0.5mm}
\noindent
\textbf{Effectiveness in sequential recommendation.}
To validate the versatility of LINK, we evaluate its performance on sequential recommendation tasks, as shown in Table~\ref{tab:exp_seqrec}. We compare LINK against state-of-the-art sequential models (SASRec, DuoRec, FEARec, and BSARec), session-based models (CORE and MSGIFSR), and the representative linear model (SLIST).
The results demonstrate LINK's capability in modeling user preferences across different evaluation scenarios, particularly for long-term sequential patterns. In sequential recommendation, user long-term histories typically involve longer average lengths and longer time intervals between items, making it crucial to capture long-term item correlations. This aligns with the results in Table~\ref{tab:exp_all}, where LINK exhibits the largest performance gains on the Tmall dataset, which has the longest session lengths. Notably, for ranking-aware metrics (\ie, MRR@20 and NDCG@20), LINK shows average performance improvements of 39.35\% and 21.53\% over BSARec~\cite{ShinCWP24}, respectively.

\begin{table*} \small
\centering
\setlength{\tabcolsep}{4.8pt} 
\caption{Performance comparison on sequential recommendation, including standard deviations for neural models. Gain indicates how much better LINK is than SLIST~\cite{ChoiKLSL21SLIST}. The best model is marked in \textbf{bold}, and the second-best model is \ul{underlined}. All experimental results for neural models are averaged over five runs using different random seeds.}\label{tab:exp_seqrec}
\vspace{-2.5mm}
\begin{tabular}{c|ccc|ccc|ccc|ccc}
\toprule
\multirow{2}{*}{Model} & \multicolumn{3}{c|}{Beauty} & \multicolumn{3}{c|}{Toys} & \multicolumn{3}{c|}{Sports} & \multicolumn{3}{c}{Yelp} \\
 & R@20 & M@20 & N@20 & R@20 & M@20 & N@20 & R@20 & M@20 & N@20 & R@20 & M@20 & N@20 \\
\midrule
CORE~\cite{hou2022core}    & 10.78\scriptsize$\pm$0.19 & 1.87\scriptsize$\pm$0.06 & 3.79\scriptsize$\pm$0.09 & 11.47\scriptsize$\pm$0.10  & 2.02\scriptsize$\pm$0.08 & 4.07\scriptsize$\pm$0.09   & 6.25\scriptsize$\pm$0.23  & 1.19\scriptsize$\pm$0.09 & 2.28\scriptsize$\pm$0.09 & \ul{11.41}\scriptsize$\pm$0.12 & 3.44\scriptsize$\pm$0.02 & \ul{5.17}\scriptsize$\pm$0.02    \\
MSGIFSR~\cite{Guo0SZWBZ22MSGIFSR} & 11.43\scriptsize$\pm$0.15 & \ul{4.34}\scriptsize$\pm$0.05 & \ul{5.89}\scriptsize$\pm$0.05 & 9.68\scriptsize$\pm$0.14  & 3.95\scriptsize$\pm$0.08 & 5.20\scriptsize$\pm$0.08   & 5.47\scriptsize$\pm$0.09  & 1.79\scriptsize$\pm$0.05 & 2.58\scriptsize$\pm$0.05 & 7.90\scriptsize$\pm$0.12   & 2.39\scriptsize$\pm$0.09 & 3.59\scriptsize$\pm$0.09\\
SASRec~\cite{KangM18SASRec}  & 11.88\scriptsize$\pm$0.06 & 3.16\scriptsize$\pm$0.05 & 5.09\scriptsize$\pm$0.05 & 12.12\scriptsize$\pm$0.09 & 3.13\scriptsize$\pm$0.06 & 5.13\scriptsize$\pm$0.05   & 6.98\scriptsize$\pm$0.12  & 1.63\scriptsize$\pm$0.06 & 2.81\scriptsize$\pm$0.06 & 8.58\scriptsize$\pm$0.14  & 3.28\scriptsize$\pm$0.03 & 4.42\scriptsize$\pm$0.05   \\
DuoRec~\cite{QiuHYW22}  & 12.21\scriptsize$\pm$0.27 & 3.45\scriptsize$\pm$0.04 & 5.37\scriptsize$\pm$0.09 & 12.83\scriptsize$\pm$0.08 & 3.62\scriptsize$\pm$0.02 & 5.67\scriptsize$\pm$0.02   & 7.29\scriptsize$\pm$0.16  & 1.90\scriptsize$\pm$0.05  & 3.08\scriptsize$\pm$0.05 & 9.21\scriptsize$\pm$0.07  & 3.33\scriptsize$\pm$0.04 & 4.60\scriptsize$\pm$0.05   \\
FEARec~\cite{DuYZQZ0LS23}  & 12.38\scriptsize$\pm$0.20  & 3.43\scriptsize$\pm$0.07 & 5.39\scriptsize$\pm$0.08 & \ul{13.06}\scriptsize$\pm$0.18 & 3.67\scriptsize$\pm$0.06 & 5.76\scriptsize$\pm$0.09   & 7.35\scriptsize$\pm$0.23  & 1.90\scriptsize$\pm$0.06  & 3.09\scriptsize$\pm$0.06 & 9.09\scriptsize$\pm$0.03  & 3.26\scriptsize$\pm$0.02 & 4.52\scriptsize$\pm$0.01   \\
BSARec~\cite{ShinCWP24}  & \textbf{12.63}\scriptsize$\pm$0.03 & 3.35\scriptsize$\pm$0.04 & 5.39\scriptsize$\pm$0.03 & 12.88\scriptsize$\pm$0.10  & 3.58\scriptsize$\pm$0.16 & 5.62\scriptsize$\pm$0.10   & \ul{7.78}\scriptsize$\pm$0.08  & 1.80\scriptsize$\pm$0.05  & 3.11\scriptsize$\pm$0.05 & 10.09\scriptsize$\pm$0.24 & \ul{3.56}\scriptsize$\pm$0.04 & 4.97\scriptsize$\pm$0.07   \\
\midrule
SLIST~\cite{ChoiKLSL21SLIST}   & 10.53   & 4.26   & 5.64   & 11.44   & \ul{5.15}    & \ul{6.55}    & 6.91   & \ul{2.65}    & \ul{3.59}    & 9.01    & 3.53   & 4.76   \\
\midrule
LINK (Ours)    & \ul{12.52}    & \textbf{4.44} & \textbf{6.20} & \textbf{13.15} & \textbf{5.46} & \textbf{7.16} & \textbf{8.17} & \textbf{2.87} & \textbf{4.03} & \textbf{11.67} & \textbf{4.01} & \textbf{5.69} \\
\midrule
Gain (\%)   & 18.94   & 4.13& 10.00      & 14.95& 6.02& 9.25& 18.21      & 8.23& 12.14      & 29.48 & 13.65      & 19.50 \\
\bottomrule
\end{tabular}
\vspace{-2mm}
\end{table*}

\subsection{In-depth Analysis}


\begin{figure*} 
\centering
\includegraphics[width=0.95\linewidth]{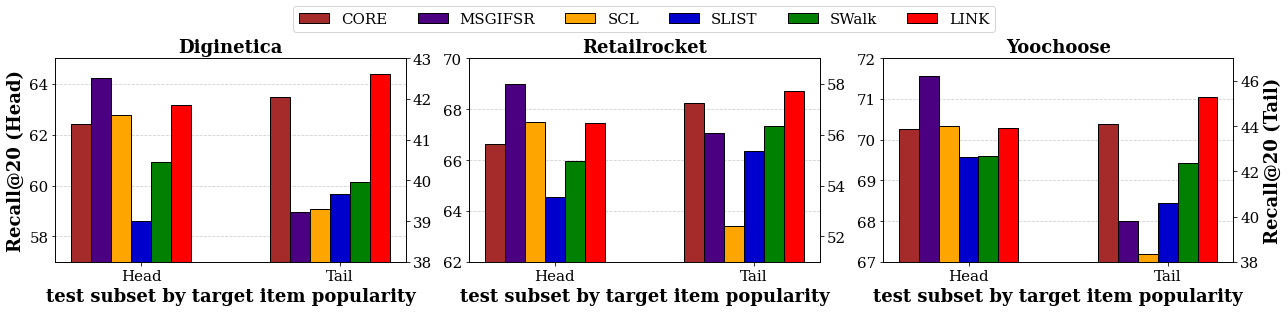}
\vspace{-2mm}
\caption{Performance comparison of representative neural and linear SBR models on Head and Tail target items. Head comprises target items within the top 20\% of popular items, while Tail consists of the bottom 80\% of less popular items.}\label{fig:exp_popularity}
\vspace{-2mm}
\end{figure*}

\vspace{0.5mm}
\noindent
\textbf{Effect of target item popularity.} We examine the ability of LINK to mitigate popularity bias by dividing the entire test set into Head and Tail, as depicted in Figure~\ref{fig:exp_popularity}. 
(i) Our analysis reveals distinct characteristics between neural and linear models. Neural models (except CORE) generally effectively capture relationships between high-popularity items, likely due to sufficient training examples. In contrast, linear models excel at modeling relationships involving low-popularity items, where their simpler architecture proves advantageous. For instance, MSGIFSR ranks 1st in Head performance while SWalk maintains robust performance on Tail items. Notably, CORE performs strongly on Tail items through its consistent representation space learning.
(ii) LINK achieves superior performance on Tail items, outperforming all baselines by up to 2.72\%. This demonstrates LINK effectively leverages linear models' capability to capture low-popularity item patterns while maintaining neural models' sophisticated relationship modeling.
(iii) Compared to the other linear models, LINK shows balanced improvements with gains of up to 3.68\% and 6.92\% on Head and Tail items. This balanced performance indicates that LINK successfully bridges the gap between linear and neural approaches.

\vspace{0.5mm}
\noindent
\textbf{Ablation study.}
To analyze the contribution of each component of LINK to overall performance, we conducted an ablation study, as shown in Table~\ref{tab:ablation}. 
(i) The two key components (\ie, LIS and NIT) improve the accuracy of R@20 by up to 2.31\% and 7.16\%, respectively. Notably, NIT demonstrates the effectiveness of distilling neural knowledge that captures complex transition patterns. When utilized together, they offer a consistent performance gain compared to a single component. 
(ii) LINK is a \textit{model-agnostic} framework that leverages the knowledge of high-capacity neural models to improve the accuracy of linear models. We employed three state-of-the-art neural models (\ie, CORE, MSGIFSR, and SCL) and observed that the performance is consistently boosted against the best linear model (\ie, SWalk). The flexibility in selecting teacher models enables LINK to adapt to diverse contexts and potentially enhance performance across various applications.
(iii) Among the teacher models, CORE proves most effective at transferring knowledge. We hypothesize that CORE's relatively simple structure, compared to GNN-based models (\ie, MSGIFSR and SCL), better aligns with linear models' architecture, enabling more efficient knowledge distillation.


\section{Conclusion}\label{sec:conclusion}

In this paper, we present the Linear Item-item model with Neural Knowledge (LINK), a novel framework that effectively combines the strengths of linear and neural models for session-based recommendation. With two components, \ie, Linear knowledge-enhanced Item-item Similarity model (LIS) and Neural knowledge-enhanced Item-item Transition model (NIT), LINK integrates the strengths of linear and neural models. Through extensive evaluation, we demonstrate that LINK successfully achieves its design goals. (i) LINK effectively captures both co-occurrence and transition patterns, leading to consistent performance improvements over existing linear models up to 14.78\%. (ii) LINK maintains competitive accuracy compared to state-of-the-art neural models while preserving the computational advantages of linear models, \ie, up to 813x fewer inference FLOPs than neural models. (iii) Model-agnostic design allows LINK to leverage advances in neural architectures. These results establish LINK as a promising approach for session-based recommendation, offering a balanced solution that combines sophisticated modeling capabilities with practical efficiency.

\begin{table}[] \small
\centering
\caption{Ablation study of LINK. The best model is marked in \textbf{bold}. Note that $\mathbf{X'}$ and $\mathcal{M}$ are in Eq.~\eqref{eq:sse_loss} and Eq.~\eqref{eq:teacher_single_item}, respectively.}\label{tab:ablation}
\vspace{-1mm}
\begin{tabular}{cc|cc|cc}
\toprule
\multicolumn{2}{c|}{Components of LINK} & \multicolumn{2}{c|}{Diginetica} & \multicolumn{2}{c}{Retailrocket} \\
$\mathbf{X'}$ for LIS & $\mathcal{M}$ for NIT & R@20 & M@20 & R@20 & M@20 \\ \midrule
\cmark & CORE~\cite{hou2022core} & \textbf{53.60} & \textbf{19.21} & \textbf{62.74} & \textbf{38.97}  \\
\cmark & MSGIFSR~\cite{Guo0SZWBZ22MSGIFSR} & 52.27 & 18.93 & 62.44 & 38.96  \\
\cmark & SCL~\cite{ShiWL24SCL} & 52.57 & 19.03 & 61.56 & 38.64  \\
\cmark & \xmark & 50.02 & 18.33 & 60.18 & 37.46 \\ \midrule
\xmark & CORE~\cite{hou2022core} & 53.57 & 19.20 & 62.66 & 38.91 \\
\xmark & MSGIFSR~\cite{Guo0SZWBZ22MSGIFSR} & 52.09 & 18.87 & 61.03 & 38.11 \\
\xmark & SCL~\cite{ShiWL24SCL} & 52.16 & 19.01 & 61.19 & 38.64  \\
\xmark & \xmark & 49.81 & 18.45 & 60.10 & 37.69 \\  \bottomrule
\end{tabular}
\vspace{-1mm}
\end{table}

\begin{acks}
    This work was partly supported by the Institute of Information \& communications Technology Planning \& evaluation (IITP) grant and the National Research Foundation of Korea (NRF) grant funded by the Korea government (MSIT) (No. RS-2019-II190421, IITP-2025-RS-2020-II201821, RS-2022-II221045, IITP-2025-RS-2024-00360227, and RS-2025-00564083, each contributing 20\% to this research).
\end{acks}

\newpage
\bibliographystyle{ACM-Reference-Format}
\bibliography{references}

\end{document}